\documentclass[twocolumn,aps,showpacs,preprintnumbers,amsmath,amssymb,ltxgrid,rotate,array,reprint]{revtex4-1}
\usepackage{graphicx,color,epsfig,tabularx,array}
\usepackage{subfig}

\begin{document}
\preprint{AIP/123-QED}

\bibliographystyle{apsrev}

\title{Probing the electric field-induced doping mechanism in YBa$_2$Cu$_3$O$_7$ using computed Cu K-edge x-ray absorption spectra}

\author{Roberta Poloni}
 \email{roberta.poloni@grenoble-inp.fr}
\affiliation{Univ. Grenoble Alpes, CNRS, SIMAP, 38000 Grenoble, France}

\author{A. Lorenzo Mariano}
\affiliation{Univ. Grenoble Alpes, CNRS, SIMAP, 38000 Grenoble, France}

\author{David Prendergast}
 \affiliation{
Molecular Foundry, Lawrence Berkeley National Laboratory, Berkeley, California 94720, USA}

\author{Javier Garcia-Barriocanal}
 \affiliation{Characterization Facility, University of Minnesota, Minneapolis 55455, USA}

\preprint{version {\today}}
\date{\today}
\begin{abstract}
We recently demonstrated that the superconductor-to-insulator transition induced by ionic liquid gating of the high temperature superconductor YBa$_2$Cu$_3$O$_7$ (YBCO) is accompanied by a deoxygenation of the sample [{\sl Perez-Munoz et al., PNAS 114, 215 (2017)}]. DFT calculations helped establish that the pronounced changes in the spectral features of the Cu K-edge absorption spectra measured in situ during the gating experiment arise from a decrease of the Cu coordination within the CuO chains. In this work, we provide a detailed analysis of the electronic structure origin of the changes in the spectra resulting from three different types of doping: i) the formation of oxygen vacancies within the CuO chains, ii) the formation of oxygen vacancies within the CuO$_2$ planes and iii) the electrostatic doping. For each case, three stoichiometries are studied and compared to the stoichiometric YBa$_2$Cu$_3$O$_7$, i.e YBa$_2$Cu$_3$O$_{6.75}$,  YBa$_2$Cu$_3$O$_{6.50}$ and  YBa$_2$Cu$_3$O$_{6.25}$. Computed vacancy formation energies further support the chain-vacancy mechanism. In the case of doping by vacancies within the chains, we study the effect of oxygen ordering on the spectral features and we clarify the connection between the polarization of the x-rays and this doping mechanism. Finally, the inclusion of the Hubbard U correction on the computed spectra for antiferromagnetic YBa$_2$Cu$_3$O$_{6.25}$ is discussed.
 \end{abstract}
\maketitle


\section{Introduction}
\begin{figure}[h]
\includegraphics[scale=0.7]{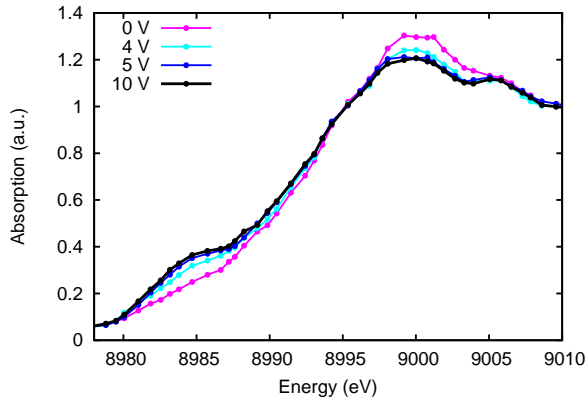}
\begin{centering}
\caption
{Experimental NEXAFS spectra measured under the applied voltage using the electric double layer technique reported in [{\sl Perez-Munoz et al., PNAS 114, 215 (2017)}].}
\label{fig1}
\end{centering}
\end{figure}

The use of electric fields to alter the conductivity of correlated
electron oxides is a powerful tool to probe their fundamental
nature as well as for the possibility of developing novel electronic
devices. The electric double layer (EDL) technique has recently been demonstrated to be an ideal tool to study the physics 
of high T$_c$ cuprates enabling controlled changes of carrier concentration by making use of an ionic liquid as a gate dielectric \cite{ShiAsa2007,YuaShi2009,YamUen2011}. 
A high density of charge carriers (as high as 10$^{15}$/cm$^{2}$) is accumulated at the oxide interface in order to screen the strong electric field generated within the EDL transistors \cite{ShiAsa2007,LeeCle2011,YeIno2010,Nakano2012}. Large doping concentration are achieved in a controlled manner allowing exploring wide regions of the phase diagram and a precise examination of boundaries between superconducting and non superconducting phases \cite{Iamada1998}. 
Recent studies of the EDL gating revealed a different doping mechanism in vanadium oxide: a large facet-dependent structural distorsion was found which was related to the migration of O atoms from the unit cell into the ionic liquid \cite{JeoAet2013,JeoAet2015}.
We recently addressed the microscopic doping mechanism responsible for the superconductor-to-insulator transition of a thin film of YBa$_2$Cu$_3$O$_7$ induced by EDL gating and collected Cu K-edge x-ray absorption spectra in situ while measuring transport properties \cite{Javipaper}.  
The measured spectra are reported in Figure \ref{fig1}. The change in the spectral features resembled those observed as a function of the O content in YBCO in early NEXAFS experiments \cite{Tranquada1988,Iwazumi1988,Tolentino1989}: the intensity increase of the low energy features together with the decrease of the white line intensity had been previously related to the increase of divalent Cu atoms within a twofold linear coordination, suggesting the possible occurrence of doping
by oxygen vacancy in our experiment \cite{Javipaper}. Also, the synthesis of YBa$_2$Cu$_3$O$_{7-\delta}$ with $\delta$=0, 0.75, 0.50 and 0.25 is known to result in geometries with oxygen vacancies localized on the CuO chain \cite{Jorge1987,Tranquada601988,Bre1988}. These observations alone could provide a plausible interpretation of the observed spectral changes. However, the doping mechanism occurring during the YBCO insulator-to-superconductor transition under EDL gating \cite{Javipaper} could have a different origin since it originates under out-of-equilibrium conditions induced by a strong electric field. For this reason, we studied three doping mechanims: i) the formation of oxygen vacancies within the CuO chains, ii) the formation of vacancies within the superconducting CuO$_2$ planes and iii) an electrostatic doping resulting from charge accumulation simulated by adding electrons to the system. 
In our previous work, we already reported spectra computed for these different doping mechanisms and established that the observed changes in the spectral features under the applied electric field could be explained by a change in Cu coordination within the CuO chains.
Here, we use DFT calculations and computed Kohn-Sham-based NEXAFS spectra to analyse the computed spectral features in terms of electronic structure changes of YBCO for each doping mechanism. In the case of doping by vacancy within the chains, we study the effect of the oxygen vacancies arrangement on the spectral features. Then, we clarify the connection between the polarization of the x-rays and the vacancy doping mechanism: as expected, the spectral changes appear evident when the the electric field direction of the x-rays has a non-zero component parallel to the direction of bond breaking of CuO along the chains. Finally, we show that although the inclusion of the Hubbard U term is critical to describe the insulating antiferromagnetic phase at the YBa$_2$Cu$_3$O$_{6.25}$ stoichiometry, the computed spectra are affected to a much lesser extent.

\section{Computational Details}
\begin{figure}[h]
\includegraphics[scale=0.3]{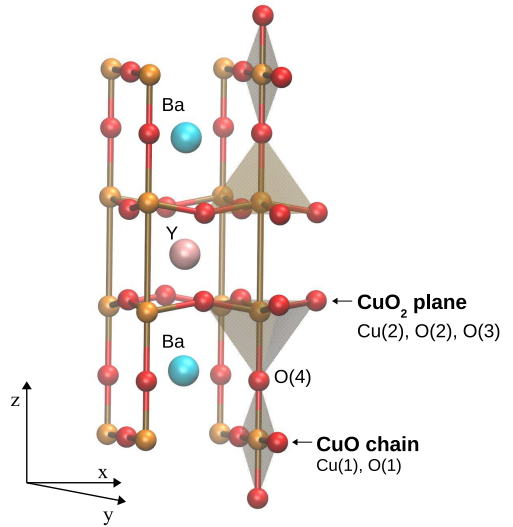}
\begin{centering}
\caption
{Crystal structure of YBa$_2$Cu$_3$O$_7$. The square pyramidal and square planar coordination geometries of the Cu atoms within the planes and chains, respectively, are highlighted for clarity. Cyan and pink spheres represent Ba and Y atoms, respectively.}
\label{fig2}
\end{centering}
\end{figure}

The electronic structure and geometrical optimization have been performed using the pwscf utility of QuantumESPRESSO \cite{quantumespresso}. We use the PBE functional and Vanderbilt ultrasoft pseudopotentials, and wavefunction and charge density cutoffs of 40 Ry and 400 Ry, respectively. Geometrical optimizations have been performed until the forces on atoms are less than 0.003 eV/\AA\ and the stress is less than 0.015 kbar. We computed the structural, electronic and NEXAFS spectra of four different stoichiometries for YBa$_2$Cu$_3$O$_{7-\delta}$, i.e. $\delta$=0, 0.25, 0.50 and 0.75. The atomic structure of orthorhombic YBa$_2$Cu$_3$O$_{7}$ is shown in Figure \ref{fig2}. For YBa$_2$Cu$_3$O$_7$, we used the 13 atoms primitive cell and a 9$\times$9$\times$3 Monkhorst-Pack grid for the intergration over reciprocal space. For vacancies created within the chains, for the YBa$_2$Cu$_3$O$_{6.50}$ composition, a 2$\times$2$\times$1 supercell with 50 atoms was used with a 5$\times$5$\times$3 Monkhorst-Pack grid. For this stoichiometry, we studied both the alternation of oxygen-full and oxygen-empty chains (see Figure \ref{fig6}c) and the checkerboard arrangement (see Figure \ref{fig6}d). Also for $\delta$=0.25, we considered both the ordered filled and empty chains arrangement (one empty CuO chain every three filled ones shown in Figure 6e) using a 4$\times$2$\times$1 supercell with 102 atoms and a checkerboard configuration with a 2$\times$2$\times$1 supercell and 51 atoms (Figure \ref{fig6}f), respectively.

For $\delta$=0.75, we considered only one configuration with an O atom vacancy surrounded in the plane of chains by vacancies, in a 2$\times$2$\times$1 supercell with 49 atoms (see Figure \ref{fig6}b). This choice is justified by experimental observations that Cu-O-Cu fragments are found to form long chains only above $\delta\sim$ 0.7. At this stoichiometry, we computed the paramagnetic metallic phase using PBE and, in order to retrieve the antiferromagnetic insulating phase, we also impose a G-type atiferromagnetic ordering using spin-polarized calculations with Hubbard-U corrections of 9 eV applied to all Cu atoms and 5 eV applied to O atoms. These values are necessary to open a gap and obtain a band structure similar to that computed by Lopez {\sl et al.} using a self interaction-error free method \cite{Lopez2010}. 
For $\delta$=0.25 and $\delta$=0.50, we considered only the paramagnetic phase in agreement with both experimental findings \cite{Kha1988} and previous computational studies \cite{Lopez2010}.

For vacancies created within the planes, we computed one atomic ordering for each stoichiometry. For $\delta$=0.50, we considered alternating filled and empty chains (Figure \ref{fig6}c) and, for $\delta$=0.25, only the ordered three-filled and one-empty chains is considered (Figure \ref{fig6}e). Since these geometries correspond to novel structures not found in previous experiments or calculations, we also computed the G-type ordering using PBE+U and found no insulating phase. Therefore, all NEXAFS spectra computed in these cases result from PBE calculations of the paramagnetic phase.

The NEXAFS spectra are computed using the fully relaxed geometries described above and, for YBa$_2$Cu$_3$O$_7$, a 2$\times$2$\times$1 supercell was used with a 3$\times$3$\times$1 Monkhorst-Pack grid. We employ the Fermi's golden rule for the calculations of the transition probabilities that enter the expression of the x-ray absorption cross section. For this, we use single-particle Kohn-Sham states for both the initial and final states. Both the electric dipole and the electric quadrupole matrix elements were computed; however, here, the latter was found to be negligible and can be safely neglected. The initial state is the 1$s$ orbital of Cu and the final states are unoccupied Kohn-Sham eigenstates derived from a self-consistent field computed within the excited electron and core-hole (XCH) approximation. The final state is computed self consistently using a core-excited pseudopotential for the excited atom and an extra electron added to the total number of ground-state valence electrons \cite{PreGal2006}. 
Within the dipole selection rule, NEXAFS accessible transitions comprise here the Cu 1$s$ hole and final states with local Cu $p$ character, the intensity of a given transition being dependent on the spatial overlap between the Cu 1$s$ orbital and the final excited state. Final states with strong $p$ character or significant localization at the excited Cu atom will lead to strong transition probabilities.
 We employed the XSPECTRA package \cite{TaiCab2002,GouCal2009} where the matrix elements are computed using the Lanczos alorithm and continued fraction for efficiency \cite{TaiCab2002,GouCal2009}. Additionally, to precisely associate the spectral features to specific electronic transitions, we employed the method developed by Prendergast {\sl et al.} in which the sum over unoccupied states is perfomed explicitly, and a Shirley interpolation scheme is employed to generate the whole Brillouin zone by using only the zone center. More details of this method can be found in Refs. \onlinecite{PreGal2006,PreLou2009}. These two methods yield similar results, however, the latter gives access to the single transition matrix elements and therefore permits identification of specific eigenstates responsible for strong spectral features. The spectra reported here are those computed using the former method.
In both cases, we adopt the same scheme for spectral alignement that allows for meaningful comparisons between different spectra computed using pseudopotentials and different periodic boundary conditions \cite{EngDuf2011}. Specifically, we compute DFT total energy differences between the total system and the isolated excited atom in the ground state and in the excited state, using the same periodic boundary conditions. This is done for each excited atom of each stoichiometry. In this way, we align the different spectra using differences in formation energies of the excited and ground state system. This procedure was not adopted when showing spectra in our previous work \cite{Javipaper}. Then, we align all spectra to the experiment by applying the same shift in the energy scale necessary to align a well-known similar material computed and measured at the same edge, here CuO.
\label{details}

\section{Crystal and electronic structure of YBa$_2$Cu$_3$O$_7$}
\label{crystal}
 In YBa$_2$Cu$_3$O$_{7}$, the Cu(1) atoms located in the CuO chains exhibit a square planar coordination in the yz plane while the Cu(2) atoms within the Cu$_2$O planes exhibit a square pyramidal coordination (see Figure \ref{fig2}). The structural and electronic evolution of bulk YBa$_2$Cu$_3$O$_{7-\delta}$ for 0$\leq$$\delta$$\leq$1 is well documented. Within a bulk material, the oxygen vacancies localize within the CuO chains and by progressively increasing their amount, the transition from the insulating antiferromagnetic phase to the superconducting one, induced by hole doping via the CuO chains, is observed \cite{Jorge1987,Tranquada601988,Bre1988}.  The change in the YBCO electronic structure around the Fermi level as a function of the oxygen stoichiometry and oxygen ordering has been reported in detail by Lopez {\sl et al.} using a self-interaction corrected method and GGA+U \cite{Lopez2010}. We refer the reader to their work for a clear and exhaustive study of the metal-insulator transition via the paramagnetic-antiferromagnetic competition in the CuO$_2$ planes and the order-disorder competition of oxygen in the CuO chains, where metallicity is related to chain formation. In this work, instead, we look at states at higher energies and in particular we focus on the unoccupied Cu $p$ states, as these correspond, here, to the dipole allowed transitions probed by the Cu K-edge NEXAFS spectra.
From the structural and electronic point of view, starting from the oxygen scarce end, i.e. YBa$_2$Cu$_3$O$_{6}$, the material is a Mott insulator exhibiting empty CuO chains (empty O(1) sites). By adding oxygen atoms in small stoichiometries (i.e. 0.75$\leq$$\delta$$\leq$1), for example in YBa$_2$Cu$_3$O$_{6.25}$, the material is still an antiferromagnetic insulator and O(1) atoms form isolated Cu-O-Cu trimers within the chains. The electronic structure exhibits a defect-like state arising from the doping O (hybridized Cu(1) $d_x^2$-O(1) $p_x$ state also localized on the apical O(4) oxygen) and lying within the Mott gap (between 0 and 1 eV in the PBE+U band structure in Figure \ref{fig8}). As the O content increases further, the trimers tend to cluster and form long chains. For stoichiometries like YBa$_2$Cu$_3$O$_{6.50}$ and YBa$_2$Cu$_3$O$_{6.75}$, alternating filled and empty-O chains are observed \cite{Kha1988} (see Figure \ref{fig7}c and \ref{fig7}e, respectively) and the electronic states localized around the O(1) atom tend to form dispersive bands across the Mott gap that overlap with the CuO$_2$ states (these bands are found between 1 and 3 eV in the case of the YBa$_2$Cu$_3$O$_{6.25}$ PBE+U band structure reported in Figure \ref{fig8}). Thus, a charge exchange mechanism between planes and chains via hybridization with the apical O(4) atoms is established, leading to the metallic state and to the hole-doped superconducting CuO$_2$ planes \cite{Lopez2010}.

\begin{figure}[ht]
\includegraphics[scale=1]{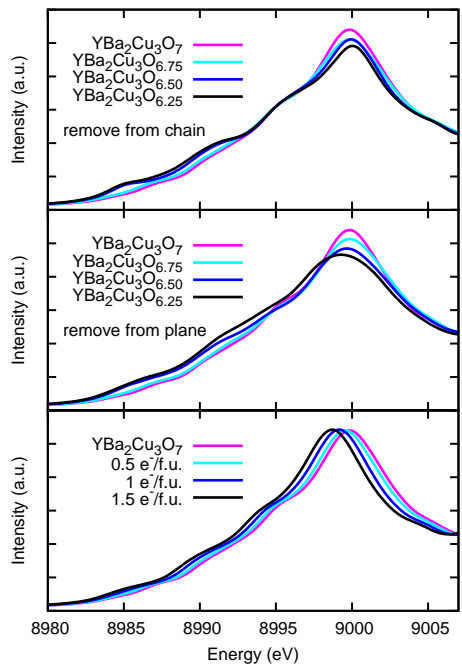}
\begin{centering}
\caption
{Computed NEXAFS spectra as a function of the oxygen content within the chain (upper panel), the planes (mid panel) and as a function of electron doping (lower panel). The spectra are computed assuming a direction of polarization of the incident beam forming 40 degrees with the diagonal of the $ab$ plane of the crystal.}
\label{fig3}
\end{centering}
\end{figure}

\section{Results}
\subsection{Doping mechanisms}
The XAS spectra measured under the external voltage are reported in Figure \ref{fig1} for an incidence angle of 40 deg between the x-ray polarization vector and the $ab$ plane of the film. An average in-plane polarization is expected in the experiment due to the presence of in-plane domains in the film, therefore, when comparing our simulations with experiments we report spectra computed with an average polarization within the ab plane and an incidence angle of 40 deg. As we will show in detail below, different spectral changes are expected depending on the alignement of the x-rays polarization with respect to the geometry of the oxygen vacancies.
 The three possible doping scenarios are simulated using a bulk approach justified by previous studies on the bulk behavior of the studied thin film \cite{LenBar2012}.  We compute i) oxygen vacancies at the CuO chain, ii) oxygen vacancies at the CuO$_2$ plane and iii) electrostatic doping.  For the electrostatic doping, we consider an increasing number of added electrons that matches the carrier concentration of the corresponding oxygen stoichiometries.
The spectra computed for the three doping mechanisms are reported in Figure \ref{fig3}. The two oxygen vacancy mechanisms (upper and medium panels) show a similar progressive increase of the intensity of the pre-edge region. The intensity of the white line also progressively decreases in both situations. However, deoxygenation from the planes exhibits a noticeable shift of the white line which is not found in the experiment. The spectra computed with electrostatic doping, instead, show a rigid shift in energy with no change in the spectral features, also not found in the experiment. As previously established, the best agreement with experiment is found when vacancies are created within the chains \cite{Javipaper}. In what follows we discuss in detail the origin of the spectral changes computed in each situation.

\subsection{Oxygen vacancies within the chains}

\begin{figure}[ht]
\includegraphics[scale=0.80]{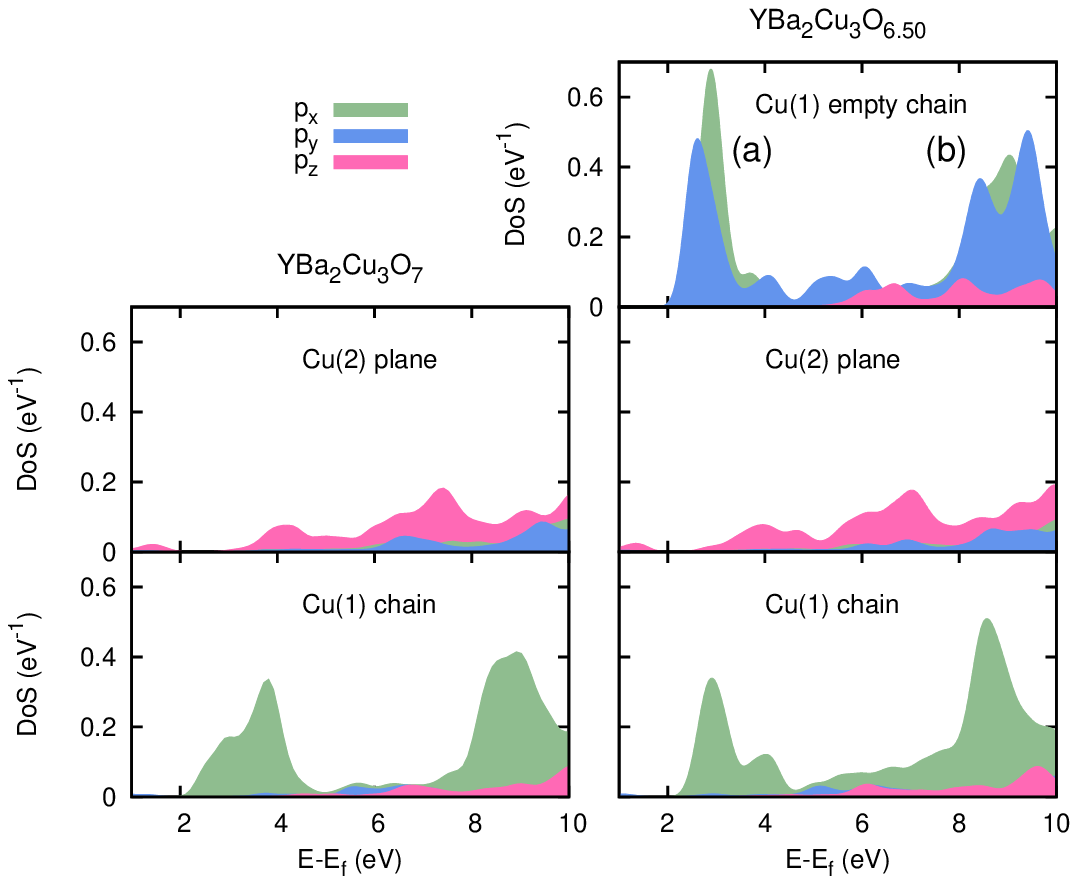}
\includegraphics[scale=0.3]{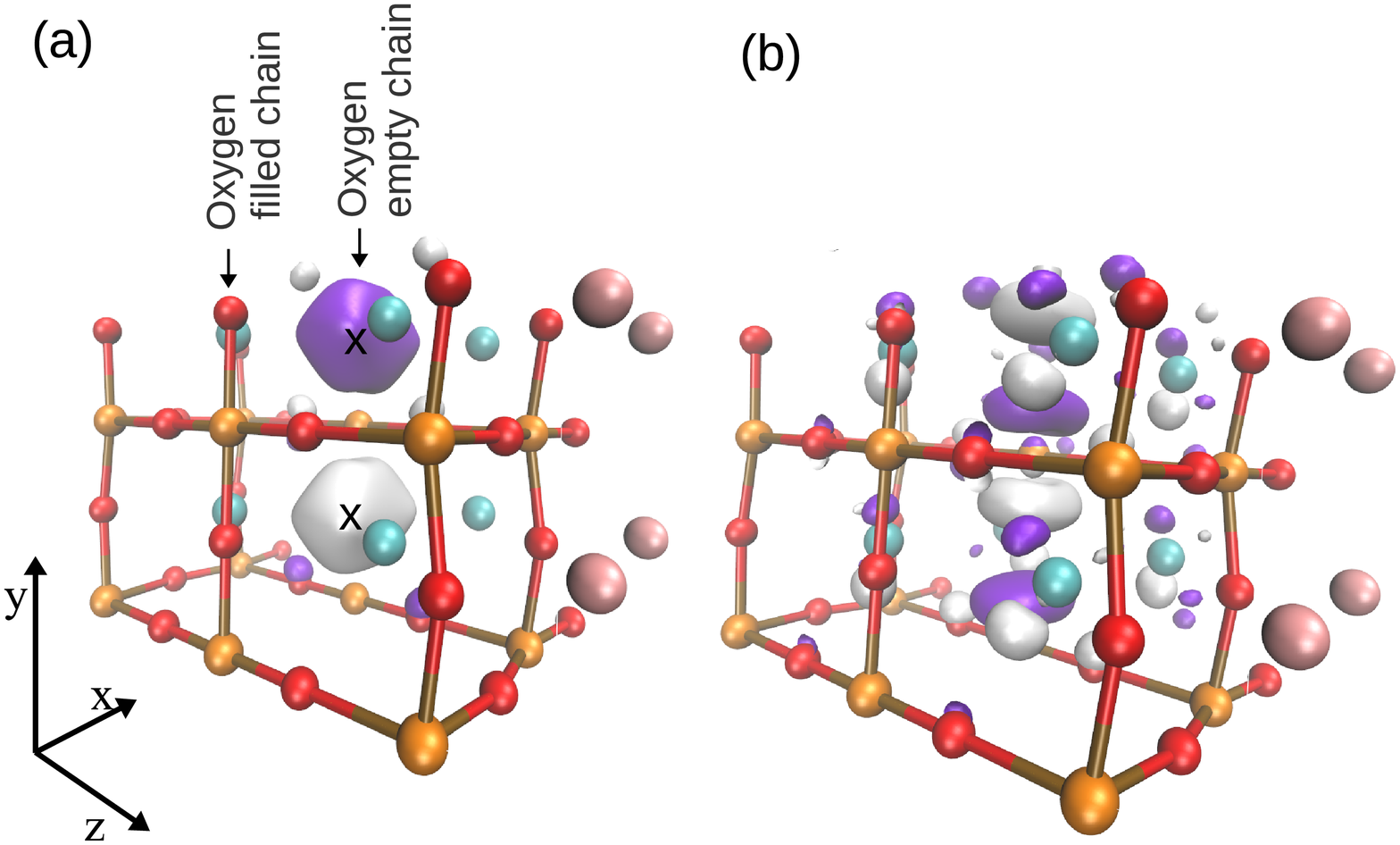}
\begin{centering}
\vspace*{-1.cm}
\caption
{Upper panels: density of states above the Fermi energy (E$_f$=0 eV) projected on the $p$ orbitals of the inequivalent Cu atoms for YBa$_2$Cu$_3$O$_{6.50}$ (left panels) and YBa$_2$Cu$_3$O$_{7}$ (right panels) with O vacancies within the chains. Lower panels: ground state orbitals corresponding to Cu(1) 4$p_y$ orbitals with bonding and antibonding character between neighboring Cu atoms along the O-empty chains. These correspond to the (a) and (b) peaks, respectively, in the density of states above (blue curves) and to the low and high energy pre-edge features in the XAS spectrum.}
\label{fig4}
\end{centering}
\end{figure}

YBa$_2$Cu$_3$O$_{7}$ exhibits two inequivalent Cu atoms as shown in Figure \ref{fig2}, Cu(1) within the chain and Cu(2) within the planes. YBa$_2$Cu$_3$O$_{6.50}$ exhibits an additional inequivalent Cu(1) atom located in the CuO chain and without neighboring oxygens along the $y$ direction. CuO chains in this work are always aligned along the $y$ direction. The change in the pre-edge region upon deoxygenation shows an intensity increase of two separate features. These features arise from unoccupied orbitals (bands) localized on the Cu(1) atoms next to O(1) vacancies.  These states are shown in the density of states above the Fermi energy projected on the $p$ orbitals of the different cupper atoms, reported in Figure \ref{fig4}.  

 For YBa$_2$Cu$_3$O$_{6.50}$, since both Cu(1) in the oxygen-empty and Cu(1) in the oxygen-filled chains have no neighboring O along $x$, their density of Cu $p_x$ states looks similar, which is also similar for Cu(1) in YBa$_2$Cu$_3$O$_{7}$ (green curves in Figure \ref{fig4}). For  Cu(1) in the empy chain, similar states are found also with $p_y$ character (upper right panel of Figure \ref{fig4}). These unoccupied non-bonding Cu 4$p_y$ states (labelled as (a) and (b) in the blue curve of Figure \ref{fig4}) arise upon deoxygenation and are due to the absense of covalent bonds with O(1).

These changes can be rationalized as follows. For YBa$_2$Cu$_3$O$_7$, Cu(1) exhibits a four-fold coordination in the $yz$ plane, and within ligand field theory, $\sigma$ metal-ligand bonds along the $y$ direction, i.e. Cu(1)-O(1) bonds, result from the overlap of Cu(1) 4$p_y$ with O(1) 2$p_y$-derived $\sigma_y$ (and from Cu(1) 3$d_{x_{^2}-y_{^2}}$ with 2$p_y$-derived states). These occupied $\sigma$ bonds are not probed here while the corresponding $\sigma^*$ antibonding orbitals are allowed electric-dipole transitions probed at the white line. The main edge corresponds also to unoccupied $\sigma^*$ states arising from the CuO$_2$ planes.  In the presence of O vacancies along $y$, the Cu(1) ligand field changes from square planar to linear, with Cu(1) atoms forming Cu-O bonds only along the $z$ direction. As a result, non-bonding Cu(1) 4$p_y$ states are probed in the pre-edge region.  As the number of vacancies increases, the number of available antibonding states from the chain decreases, reducing the transition probabilities. Accordingly, spectral weight is transferd from the main edge to the pre-edge as shown in Figure \ref{fig3}. 
\begin{figure}[h]
\begin{centering}
\includegraphics[scale=1.]{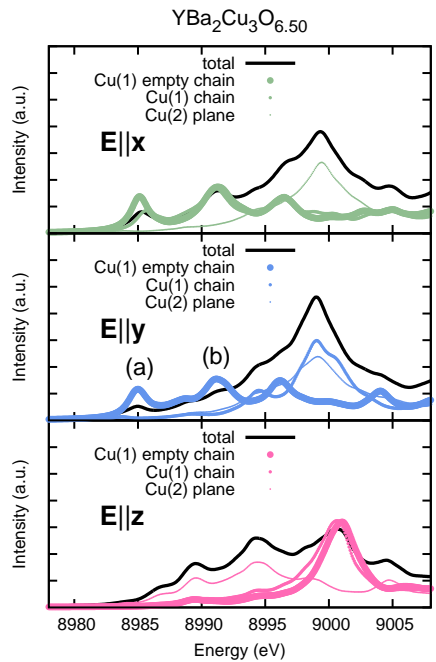}
\caption
{The spectrum of YBa$_2$Cu$_3$O$_{6.50}$ computed using a polarization of the x-ray parallel to $x$, $y$ and $z$ direction of the unit cell is shown in the upper, middle and lower panels, respectively. The contribution from each inequivalent atom is also shown for clarity.}
\label{fig5}
\end{centering}
\end{figure}
The low energy spectral features (arising from (a) at 8985 eV and (b) at 8990 eV in Figure \ref{fig4}) correspond to non-bonding 4$p_y$ states forming localized bands along $y$. Both of these states are non-bonding in the sense of non covalent-bonding, but the symmetry in the phase of the orbital at $\Gamma$ along adjacent Cu atoms in the empty chain permits a bonding (a) and anti-bonding (b) distinction. The ground state wavefunction at $\Gamma$ corresponding to these two transitions is shown in the bottom panels of Figure \ref{fig4}. We plot here for clarity the ground state orbital rather than the excited state orbital. The excited state orbital is similar but significantly more localized on the excited atom.

In the upper, middle and lower panels of Figure \ref{fig5}, we show the spectra of YBa$_2$Cu$_3$O$_{6.50}$ computed with the polarization of the electric field aligned along the $x$, $y$ and $z$ direction of the YBCO, respectively. For clarity, the contribution from each inequivalent Cu atom to the total spectrum is also reported. 
\begin{figure}[h]
\begin{centering}
\includegraphics[scale=0.33]{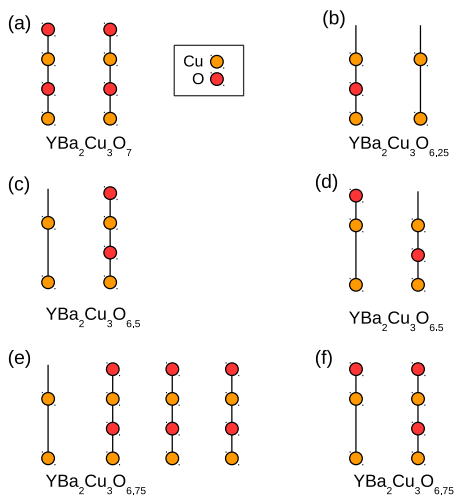}
\includegraphics[scale=1]{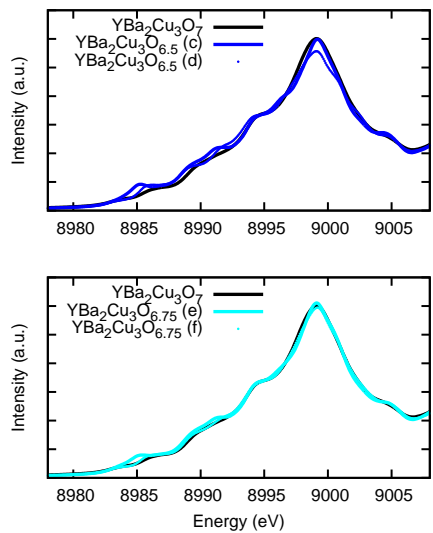}
\caption
{Upper panel: oxygen vacancy arrangements considered in this work; YBa$_2$Cu$_3$O$_{7}$ is shown in (a) as a reference; (b) YBa$_2$Cu$_3$O$_{6.25}$; (c) and (d) show the alternating empty and filled chains and checkerboard orderings for YBa$_2$Cu$_3$O$_{6.25}$; (e) and (f) show the aternating two filled and one empty chains and checkerboard orderings for YBa$_2$Cu$_3$O$_{6.75}$. Lower panel: spectra computed for the vacancy arrangements described above. More pronounced spectral changes are predicted for alternating empty and filled chains.}
\label{fig6}
\end{centering}
\end{figure}
As expected, for polarization along $x$, the contribution from Cu(1) in the O-empty and O-filled chains is similar, meaning that the creation of an oxygen vacancy would negligibly modify the spectrum in this case. When the electric field is polarized along $y$, the non-bonding Cu(1) 4$p_y$ states would lead to allowed transitions with signficant spectral weight. The spectral contribution from these states, labelled as (a) and (b) in Figure \ref{fig4}, are also shown in the middle panel of Figure \ref{fig5}, for clarity. For a polarization along $z$, a negligible change is expected, as in the $x$ case. Thus, the spectral changes are detectable when the direction of the polarization of the x-rays has a non-zero component parallel to $y$ axis of the sample.

Next, we studied the two different vacancy orderings for YBa$_2$Cu$_3$O$_{6.50}$ and YBa$_2$Cu$_3$O$_{6.75}$. The spectra reported in Figure \ref{fig3} correspond to alternating O-empty and O-filled long chains of vacancies along $y$ \cite{Kha1988}, i.e., an alternation of one O-empty and one O-filled chains (Figure \ref{fig6}c) in the first case and two O-filled chains followed by an O-empty chain in the second  (Figure \ref{fig6}e). The comparison between O-empty and O-filled chains with checkerboard geometries for each stoichiometry is shown in the lower panels of Figure \ref{fig6}, together with the spectrum of pristine YBa$_2$Cu$_3$O$_{7}$. We find similar spectral changes at the pre-edge and white line for the two arrangements but more pronounced changes are noticeable for the alternation of empty and filled chains.

\subsection{Oxygen vacancies within the planes}
\begin{figure}[ht]
\includegraphics[scale=1.]{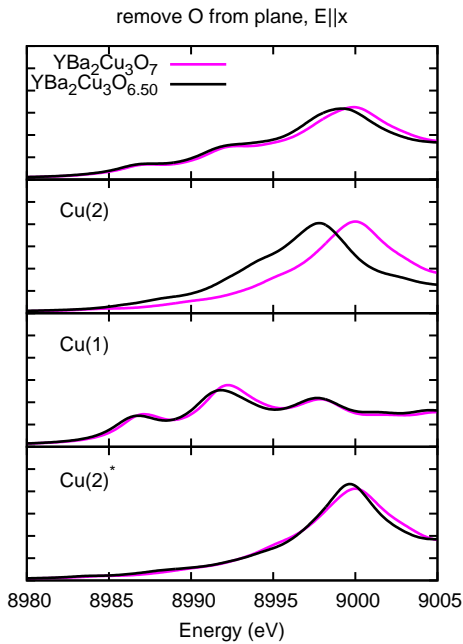}
\begin{centering}
\caption
{Atom-decomposed spectra for YBa$_2$Cu$_3$O$_7$ and YBa$_2$Cu$_3$O$_{6.50}$ with vacancies along the $y$ direction within one of the two CuO$_2$ planes. Only the spectra computed for the electric field vector of the electromagnetic field parallel to the $x$ direction are considered here for clarity.}
\label{fig7}
\end{centering}
\end{figure}

The middle panel of Figure \ref{fig3} shows that for O vacancies that are created within the CuO$_2$ planes, the two low energy features appear, in accordance with the change in ligand field of the decoordinated Cu atoms. In this case, these features derive from non-bonding Cu(2) 4$p$ states along $y$, located within one CuO$_2$ plane.  A clear redshift of the whole spectrum is also noticeable, contrary to the previous case and to the experimental observations.  The analysis of the crystal and electronic structure of the deoxygenated compounds reveals a chemical origin of this shift. For YBa$_2$Cu$_3$O$_{6.50}$, i.e. upon removal of two O atoms along $y$ within one plane, three inequivalent atoms are formed: the Cu(2) atoms that keep a five-fold coordination in the plane where two O are removed, the Cu(2)$^*$ atoms that are now three-fold coordinated and the Cu(1) atoms on the chains. In reality, there are more than three inequivalent atoms since the second plane also modifies its atomic arrangement. However, we have found negligible differences in the corresponding spectra. The atom-decomposed spectra of YBa$_2$Cu$_3$O$_{6.50}$ shown in Figure \ref{fig7} reveals that the origin of the shift arises from the $x$ component of the polarization. Atoms Cu(2) and Cu(2)$^*$ exhibit a main peak corresponding to the white line of the full spectrum. Since both Cu(2) and Cu(2)$^*$ atoms form covalent bonds along $x$, the antibonding $\sigma^*$ states provide the most intense dipole-allowed transitions in both cases. Cu(1) atoms instead have no bonding along $x$  and transitions to the non bonding Cu(1) 4$p_x$ states are found.  The third feature at around 8998 eV also represents a transition to a non bonding Cu 4$p_x$ band with antibonding character between neighboring Cu atoms and with a significant contribution from all neighboring Ba states.
 
We notice that the structure of YBa$_2$Cu$_3$O$_{6.50}$ exhibits a signficant modification of the CuO bond lengths within the vacancy-doped plane. Shorter Cu(2)-O(3) bond lengths along $x$ are found for O(2) vacancies along the $y$ direction. This results in alternating chains (within the plane) of Cu atoms and CuO bonds along $y$ whit CuO bond lengths along $x$ of 1.86 \AA\ and 2.06 \AA\ (as a reference, the bond length of planar CuO in YBa$_2$Cu$_3$O$_7$ is 1.92 \AA), respectively.  Our analysis of the projected density of states indicates that the longer bonds of the Cu(2) atoms exhibits weaker bonding and lower antibonding states compared to YBa$_2$Cu$_3$O$_7$, resulting in spectra peaked at lower energy, as illustrated in the second lower panel of Figure \ref{fig7}.

  We also find a slightly larger charge on the atoms with shorter bonds, resulting from the absence of $\pi$ backbonding donation. However, this feature contributes negligibly to a shift of the spectra as illustrated in the bottom panel of Figure 6.
When three vacancies are created within the plane leading to a stoichiometry of YBa$_2$Cu$_3$O$_{6.25}$, the situation is similar, with CuO bonds along $x$ of 1.91 and 2.00 \AA\ for the Cu atoms without bonding along $y$ and with a single bond along $y$, respectively. The bond length change is smaller now but more Cu atoms contribute to the shift resulting in a more noticeable change of the whole spectrum.


\subsection{Oxygen vacancy formation energy}
We computed the O vacancy formation energy for five configurations. For YBa$_2$Cu$_3$O$_{6.75}$ in configuration (c) of Figure \ref{fig6}, i.e. one vacancy surrounded by O atoms, we computed the formation energy of a vacancy i) within the chain and ii) within the plane. These configurations are fair approximations to isolated vacancies. Then, for YBa$_2$Cu$_3$O$_{6.50}$ we computed the formation energy of two vacancies iii) within the chain in configuration (c) of Figure \ref{fig6}, iv) in configuration (d) and v) O vacancies in the plane in configuration (c).  By using total energies only and therefore neglecting thermal and vibrational terms, the O vacancy formation energy can be computed as follows:
\begin{equation}\label{eq1}
\Delta E_{vac} = (E_{vac}-E_{stoi}-\frac{1}{2}E_{O_2}),
\end{equation}
where E$_{vac}$ is the total energy of YBa$_2$Cu$_3$O$_{7-\delta}$ and E$_{stoi}$ is total energy of stoichiometric YBCO, i.e. YBa$_2$Cu$_3$O$_7$. The computed $\Delta E_{vac}$ for YBa$_2$Cu$_3$O$_{6.75}$  in the i) chain and ii) in the plane are 1.40 and 2.30 eV, respectively, showing that an isolated vacancy would be preferably located within the chain. Then, for YBa$_2$Cu$_3$O$_{6.50}$ the computed $\Delta E_{vac}$ per O vacancy are 1.20 eV, 1.46 eV and 2.10 eV for iii), iv) and v), respectively, showing that two vacancies prefer to align on the same chain rather than in a checkerboard configuration. Again, even when vacancies align within the plane, they are thermodynamically less favorable then any computed chain configuration. Thus, for a bulk system and in absence of electric field, once an O vacancy is created in the chain, additional vacancies tend to localize next to it within the same chain.

\subsection{Electrostatic doping}
The lower panel of Figure 3 shows the spectra computed for YBa$_2$Cu$_3$O$_7$ with an addition of 0.5, 1 and 1.5 electrons per formula unit to simulate the electrostatic doping corresponding to the amount of carriers added by doping with vacancies. We observe an almost-rigid redshift of the spectrum consistent with the increase in the Fermi energy, e.g. between the 0.5 electron and 1.5 electrons calculations, the 0.78 eV difference in Fermi energy well accounts for the computed shift in the corresponding spectra. Additional electrons result in an increased occupation of bands as evidenced by the density of states near the Fermi energy with minor changes in the shape of the $p$-projected density of unoccupied states. This can be interpreted as an electrostatic effect with the redshift resulting from a more weakly bound core electron. The almost-rigid shift also suggests a nearly uniform localization of the additional electrons on the Cu atoms. This is indeed the case, and the electronic charge localizes only 10\% more on the Cu atoms of the chain, as compared to those in the planes, due to the higher electronegativity.

\subsection{Spectrum computed with PBE+U}
\label{PBE+U} 
\begin{figure}[h]
\includegraphics[scale=1]{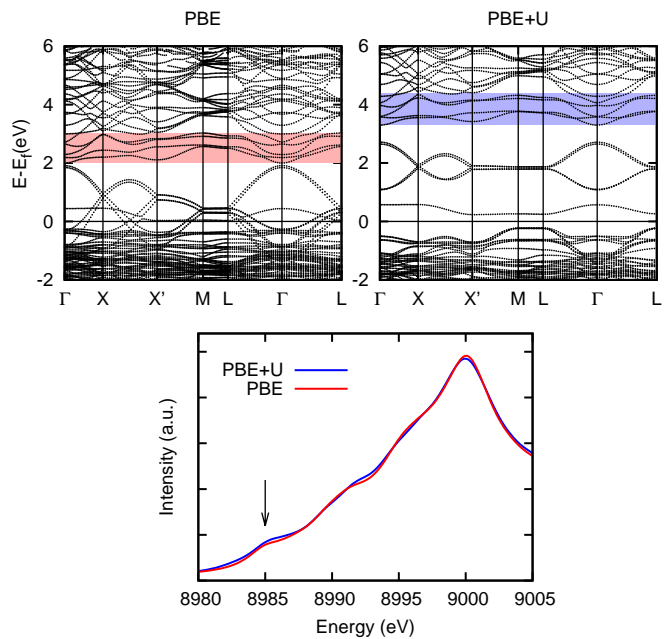}
\begin{centering}
\caption
{Electronic band structure of YBa$_2$Cu$_3$O$_{6.25}$ computed with PBE (upper left panel) and PBE+U (upper right panel). The red and blue rectangles show the first unoccupied Cu states with $p$ character, the Cu(1) 4$p_x$ and 4$p_y$ bands. Transitions of the core electron to these states are shown with an arrow. In the PBE+U calculations these states are located about 1.4 eV higher in energy with respect to the Fermi level and we use this value to align the corresponding spectrum to the PBE one.}
\label{fig8}
\end{centering}
\end{figure}

As discussed above, for YBa$_2$Cu$_3$O$_{6.25}$, an insulating antiferromagnetic phase is expected. However, even if we impose an antiferromagnetic G-type ordering using PBE, a metallic state is obtained. In order to open a gap, the self-interaction error intrinsic to standard functionals which mostly affects localized orbitals should be corrected. Here, we show, in agreement with previous results by Lopez {\sl et al.} \cite{Lopez2010}, that a Hubbard-U correction of 9 eV needs to be applied to the Cu $d$ orbitals. This value was reported in Ref. \onlinecite{Lopez2010} and was found to give a similar band structure to a SI-free DFT calculation reported in the same work.  In addition, we have found that a 5 eV correction should be applied also to the O $p$ states that hybridize with the planar Cu atoms, in order to predict the antiferromagnetic insulating phase. Figure \ref{fig8} reports the electronic band structure computed using PBE and PBE+U and the corresponding NEXAFS spectra. Although the states near the Fermi energy are largely modified, the two spectra are relatively similar with all important features reproduced similarly. The most noticeable difference is found in the intensity of the pre-edge features. These features correspond to transitions to the bands located between 2 and 3 eV in the PBE calculation and between 3.4 and and 4.5 eV in the PBE+U calculation. These are the Cu(1) 4$p_x$ and 4$p_y$ states discussed above. Our analysis reveals that when PBE+U is used, more localized bands are found, resulting in larger transition probabilities. As discussed above, the localized bands located right above the Fermi energy and below the Cu 4$p$ states of the PBE+U band structure are spin-degenerate $\sigma^*$ Cu 3$d$-O 2$p$ states localized on the trimers (between 0 and 1 eV) and $\sigma^*$ Cu(2) $d_{x^2-y^2}$-O(2,3) $p_{x,y}$-derived states (between 1 and 3 eV). Within dipole-selection rule and in absence of significant quadrupole contribution and/or spin-orbit coupling, transitions to these states are symmetry-forbidden.

Figure \ref{fig8} shows that the first allowed transitions in the PBE+U calculations are located 1.4 eV higher with respect to the Fermi energy, as a result of the gap opening. We have thefore applied this shift in the energy axis to the PBE+U spectrum for the purpose of alignement with the PBE one, and we found that all higher energy available states are moved up (scissor-shift) by the same quantity. In conclusion, the two spectra are similar but not identical with high-energy states are negligibly affected by the inclusion of Hubbard-U and low-lying states are only slightly affected. This is in contrast with previous calculations of Ni K-edge spectra computed for NiO for which a significant pre-edge spectral change upon inclusion of the Hubbard-U was predicted due to the strong offsite mixing of Ni 4$p$ states with the Ni 3$d$/O 2$p$ upper Hubbard band \cite{GouCal2009}. Here the situation is different since the pre-edge features belong to localized Cu 4$p$ bands onto which the Hubbard-U has a minor effect compared to $d$ states, consistent with the large separation between Cu 3$d$ bands right above the Fermi level and the higher energy Cu 4$p$ bands (no mixing between these states).  

\section{Conclusion}
We report NEXAFS spectra  computed for three different doping mechanisms of bulk YBCO: i) doping by O vacancies within the chain, ii) doping by O vacancies within the planes and iii) electrostatic doping. A purely electrostatic doping can be safely ruled out as the computed almost-rigid shift of the spectra does not correspond to the experimental observations. The spectral changes computed for O vacancies within the CuO chains best resemble the experimental data: a progressive increase in the intensity of the pre-edge features and decrease in the intensity in the white line is predicted and experimentally detected. When O vacancies are created within the chain, the Cu coordination of these Cu atoms change from four-fold to linear, resulting in the formation of non bonding 4p states (probed at the pre-edge) between neighboring Cu atoms along the empty chain and in the absense of antibonding CuO states along the chain (probed at the white line). O vacancies within the CuO$_2$ plane result in spectral features similar to the chain-vacancy case, however a clear red-shift of the white line is predicted but not measured. 
In the case of O vacancies located within the chain, we analyze the atom-decomposed contribution for better interpretation, the sensitivity of the spectral changes on the vacancy arrangements and on the orientation of the sample with respect to the polarization of the x-rays.   
Differencies in the computed vacancy formation energies also support the preferred chain-vacancy picture. This study suggests the electrochemical doping occurring upon EDL gating and reported in Ref. \onlinecite{Javipaper} likely results in deoxygenated YBCO samples simlar to those obtained with standard synthesis methods.

\begin{acknowledgments}
Calculations were performed using resources granted by GENCI under the CINES grant number A0020907211 and A0040907211. Additionally, the froggy platform of the CIMENT infrastructure was employed. D.P.'s work at The Molecular Foundry is supported by the Office of Science, Office of Basic Energy Sciences, of the United States Department of Energy under Contract No. DE-AC02-05CH11231.
\end{acknowledgments}

\onecolumngrid

\end{document}